\documentclass[aps,pre,reprint,twocolumn,superscriptaddress,showpacs,nofootinbib,longbibliography]{revtex4-2}
\usepackage{graphicx,amsbsy,amsmath,epsfig,natbib,float}
\usepackage{amssymb,latexsym,wasysym,pifont,mathrsfs}
\usepackage{comment,verbatim,multirow,array,sidecap,color,bm}
\usepackage[normalem]{ulem}
\usepackage{hyperref}

\definecolor{blueberry}{rgb}{0.01569, 0.2, 1.0}
\definecolor{asparagus}{rgb}{0.4,0.4,0}
\definecolor{maraschino}{rgb}{1.0,0.0,0.0}
\definecolor{strawberry}{rgb}{1.0,0.0,0.5}

\definecolor{mygreen}{rgb}{0.0,0.55,0.3}
\definecolor{maroon}{rgb}{0.7,0.0,0.2}
\definecolor{teal}{rgb}{0.0,0.5,0.5}

\begin{document}

\title{Cyclically sheared colloidal gels: structural change and delayed failure time}

\author{Himangsu Bhaumik}
\affiliation{Yusuf Hamied Department of Chemistry, University of Cambridge, Lensfield Road, Cambridge CB2 1EW, UK}

\author{James E. Hallett}
\affiliation{Department of Chemistry, School of Chemistry, Food and Pharmacy, University of Reading, Reading RG6 6AD, United Kingdom}

\author{Tanniemola B. Liverpool}
\affiliation{School of Mathematics, University of Bristol, Fry Building, Bristol BS8 1UG, UK}

\author{Robert L. Jack}
\affiliation{Yusuf Hamied Department of Chemistry, University of Cambridge, Lensfield Road, Cambridge CB2 1EW, UK}
\affiliation{DAMTP, Centre for Mathematical Sciences, University of Cambridge, Wilberforce Road, Cambridge CB3 0WA, UK}

\author{C. Patrick Royall}
\affiliation{Gulliver UMR CNRS 7083, ESPCI Paris, Université PSL, 75005 Paris, France}

\date{\today}
 
\begin{abstract}  
We present experiments and simulations on cyclically sheared colloidal gels, and probe their behaviour on several different length scales.  The shearing induces structural changes in the experimental gel, changing particles' neighborhoods and reorganizing the mesoscopic pores.  These results are mirrored in computer simulations of a model gel-former, which show how the material evolves down the energy landscape under shearing, for small strains. By systematic variation of simulation parameters, we characterise the structural and mechanical changes that take place under shear, including both yielding and strain-hardening.   
We simulate creeping flow under constant shear stress, for gels that were previously subject to cyclic shear, showing that strain-hardening also increases gel stability.  This response depends on the orientation of the applied shear stress, revealing that the cyclic shear imprints anisotropic structural features into the gel.
\end{abstract}

\maketitle 

\newcommand{\eps}{\varepsilon}
\newcommand{\eprep}{\eps_{0,\rm prep}}
\newcommand{\gmax}{\gamma_{\rm max}}
\newcommand{\ncyc}{n_{\rm cyc}}

\newcommand{\tausim}{\tau_{\rm sim}}

\section{Introduction}

Colloidal gels are assemblies of particles consisting of interconnected networks of strands, which are kinetically arrested far from equilibrium~\cite{zaccarelli2007colloidal,
royall2021real,cipelletti2005slow,poon2002physics,lu2008gelation,lu2013}.  They exhibit complex relaxations and responses to external forces, whose understanding is important for applications~\cite{zaccarelli2007colloidal,
royall2021real}, and as a fundamental outstanding problem in soft matter modelling~\cite{segre2001glasslike,sedgwick2005non,landrum2016delayed,tsurusawa2019direct,rouwhorst2020nonequilibrium}.  These materials are particularly challenging because their behaviour is influenced by structural features on different length scales, including local motifs in the microscopic structure~\cite{patrick2008direct}, the response of individual strands to applied stress~\cite{van2018strand,verweij2019plasticity,KrisSMNeck23}, and the heterogeneous network~\cite{WeitzPRL06, HsiaoPNAS12, bouzid2017elastically,JamaliPNAS24,JamaliPRE24}.

Gel structure depends on particle interactions (strength and range of attractive forces) and on their volume fraction.  In addition, the fact that gels are far from equilibrium means that their structures also depend on their mechanical and thermal history; they also experience physical aging, so their properties depend on the  time elapsed since their preparation~\cite{fielding2000aging,trappe2001jamming,WeitzPRL2000Universal, zia2014micro,nabizadeh2021life,bartlett2012sudden, patinet2016connecting, parley2020aging, pollard2022yielding}.  This feature can be exploited to engineer materials with specific properties, but the relationships between history, structure and gel properties are complex, and theoretical predictions are limited, so that formulation of gels often requires a large component of trial-and-error.

Among the gel properties that one would like to control are the linear response to external stress (compliance) and the yielding behavior.  The process of strain-hardening offers a promising route towards this control, in that mechanical processing of an already-formulated material can be used to suppress yielding and/or reduce compliance. 
Strain-hardening is familiar in other amorphous solids such as {molecular and metallic} glasses~\cite{Lee2009science,Deng2012,Leishangthem2017} where the shear provides a mechanism for the system to descend in its energy landscape and form more stable structures (see also~\cite{Kim2024nature}).
There is also experimental evidence for strain-hardening in the rheology of fractal colloidal gels~\cite{Gisler1999} (see also~\cite{Pouzot2006,Schmoller2010,Reis2019}),
although bulk rheological measurements cannot characterise the microscopic changes in gel structure that underlie the hardening (and thinning) processes. 
The network structure of a gel points to a more complex rheological response than glasses.  In particular, disruption of the network by shearing is a mechanism for thixotropy (shear-thinning)~\cite{BonnRMP2017} see also~\cite{colombo2014stress,Bouzid2018langmuir,lockwood2024}; shear-thinning can also occur in glasses, but the underlying physical mechanism is different~\cite{Mizuno2024}.

This work reports experiments and computer simulations of gels that form by depletion in colloid-polymer mixtures.  The experiments
combine a shear stage with \emph{in situ} particle-resolved imaging by 3d confocal microscopy, enabling microscopic changes in structure to be probed.
Such methods 
have extensively been used to investigate colloidal glasses~\cite{Royall2024,Besseling2007,Schall2007}. 
In the case of colloidal gels, combined imaging and shear enables direct observation of shear induced microstructural changes~\cite{ballesta2013,conrad2008} and 
has been shown to produce mechanically strong gels by rejuvenation~\cite{koumakis2015tuning}, see also~\cite{moghimi2017residual,Rajaram2010}.

In previous work using rheological methods, start-up shear has revealed two-step failure in gels, at least in the range of volume fraction of interest here~\cite{KoumakisSM11}. Pioneering work with cyclic shear on fractal gels of polystyrene particles at quite low volume fraction ($\phi<0.05$) found that strain {hardening} was possible~\cite{Gisler1999}. More recent studies probed fatigue in carbon black gels, where the interparticle bonds are typically rather stronger than those we consider here~\cite{perge2014}. In weaker gels, large amplitude oscillatory shear has been shown to weaken and break the network, but small amplitudes strengthen the material~\cite{moghimi2017colloidal}. Recently, a gel with van der Waals interactions was subjected to cyclic shear, which imprinted a memory in the network, and led to strengthening~\cite{CohenSMMemory2020} (this effect is isotropic, even though the shear breaks rotational symmetry). 

Our experiments consider the effects of a cyclic shear protocol on gel structure and properties. 
Strong shearing can exceed the yield stress and cause the gel to break; it is also possible that damage can accumulate over many shear cycles, leading to fatigue failure~\cite{maity2024arxiv}.   As well as these effects -- which weaken the material -- we also find strain-hardening effects in some cases.  
The structural evolution involves two important processes: the breaking of gel strands under shear~\cite{van2018strand,verweij2019plasticity,KrisSMNeck23}, and the evolution of the microscopic structure towards lower energies~\cite{bhaumik2021role}.
These are evidenced by  pore-size distributions (mesoscopic structure) and the topological cluster classification (TCC)~\cite{malins2013tcc} for the (microscopic structure).  The same behavior is also present in our computer simulations.

A second part of our study concerns the dynamical behavior of simulated gels, including the rheological properties during cyclic shear, and measurements of the yielding under constant stress (creep).  We present evidence for strain hardening, based on compliance and dissipation during the shear cycles.  We also subject the resulting gels to constant stress, which leads to creeping flow and eventual yielding.  These samples survive for longer times before failure, compared to gels that were not strain-hardened.
 Moreover, this enhanced stability depends significantly on the relative directions of the shearing motions for the cyclic and creeping protocols.   In other words, the hardening can be used to enhance gel stability and to imprint anisotropic responses to subsequent shearing.

These results demonstrate specific situations in which gel properties can be selected by mechanical processing.  The particle-resolved experiments reveal the microscopic and mesoscopic changes in structure; the simulations show how these can be harnessed to enhance gel stability and to tailor  anisotropic responses.  We discuss how our understanding of these far from equilibrium materials might be harnessed to for prediction and design of  material properties more generally.

The structure of the paper is as follows: Sec.~\ref{sec:methods} describes models and methods and Sec.~\ref{sec:results-structure} presents results for structural change due to cyclic shear.  Sec.~\ref{sec:results-rheo} discusses the dynamical behaviour, after which Sec.~\ref{sec:conc} concludes with a discussion of our results and directions for future work.

\section{Models and Methods}
\label{sec:methods}

\subsection{Experimental}

\begin{table}
 \begin{tabular}{|l|c|c|c|}
\hline
$\eps$ & $\dot\gamma$ ($/\tau_B^{-1}$)  & $\gmax$ & $\phi$ \\
\hline
{$c_p = 1.25 c_p^*$}  & \; 3.72 \, & \,0.0337\, & \,0.17\, \\
{$c_p = 1.50 c_p^*$} &  9.32  & \,0.0337\, & \,0.18\, \\
\hline
 \end{tabular}
    \caption{Experimental parameters for shear.  The rate $\dot\gamma$ is given relative to the Brownian time.  
 }
\label{tab:shear}
\end{table}

We consider sterically stablised {poly-methyl methacrylate} 
colloids in a cis-decalin and cyclohexylbromide solvent mixture which matches the density and refractive index.  They are dispersed in a 
solution of (4mM) of tetrabutyl ammonium bromide to screen the electrostatic interactions. Depletion attractions are provided by a nonadsorbing polymer (polystyrene, molecular weight 8.4MDa).   The (mean) particle diameter is $\ell=1500\rm nm$, we estimate the polydispersity at 5\%.  The colloid volume fraction is  {given in Tab.~\ref{tab:shear}.}
We estimate the Brownian time as $\tau_B=4.0$s.
   
We present data for polymer concentration in the range from $3.42-4.10$gl$^{-1}$, see also below.  
We estimate the polymer radius of gyration $R_g$in this solvent as $155$nm  by matching to a one-component effective Asakura-Oosawa potential and using the Noro-Frenkel criterion~\cite{noro2000} to match interaction strength at the onset of gelation (at concentration 2.73gl$^{-1}$). This gives a polymer-colloid size ratio $2R_g/\ell\approx 0.207$.

 The sample dimensions are $1$cm$\times 1$cm$\times100\mu$m.
 We extract the positions of the particles from confocal microscopy images using the trackpy package. 
 We consider the particles inside a cubic visualisation region (``box'')
of linear size $L\approx 33\ell$. 
This results in co-ordinates of $N\approx 1.5\times 10^4$ particles.

We use a shear stage to apply cyclic (simple) shear with motion in the $y$ direction and velocity gradient in $z$ direction. The shear stage comprised an aluminium flexure with a coverslip window, mounted on a plexiglass baseplate and separated by a rubber spacer, which served to both dictate the gap thickness and to prevent solvent evaporation. Both the top coverslip and bottom glass surface were coated with a sintered particle layer to allow the gel to adhere to both surfaces. Displacement of the top window was driven by a Thorlabs TPZ001 T-cube piezo driver connected to a Thorlabs piezo stack with free stroke displacement of 25.5 microns, controlled by a MATLAB script. The shear is a triangular wave with constant rate $\dot\gamma$, that is, the strain is increased linearly in time from $\gamma=0$ to $\gamma=\gamma_{\rm max}$ and then decreased to $\gamma=-\gamma_{\rm max}$ before finally increasing back to $\gamma=0$. We alternate shearing and imaging, the acquisition time for an $3d$ image of the gel is approximately $30$s ($7.5\tau_B$); there is negligible aging during this time.   
The location of the visualisation region does not stay constant with time, this ensures that data is not too much affected by photobleaching that would occur if all imaging occurred in the same place. The waiting time $t_{\rm w}$ between sample loading and the start of shearing is approximately 30 minutes ($450\tau_B$).

We consider two state points that differ in the polymer concentration and the parameters of the applied shear.  We write $\varepsilon^*$ for the interaction strength at the onset of gelation.  The interaction strengths for our reported results are $\eps=1.25\eps^*,1.50\eps^*$.   The parameters for the shear {experiments} are reported in Table~\ref{tab:shear}.

\subsection{Computational}

\subsubsection{Model}

{For numerical simulations, we use an established model of a size-polydisperse colloid-polymer mixture~\cite{patrick2008direct, taffs2010, zia2014micro, landrum2016delayed, razali2017effects, Griffiths2017}
which has been accurately mapped to the standard two-component Asakura-Oosawa model~\cite{asakura1954,taffs2010}.   The depletion interaction between colloidal particles is modeled by a} truncated and shifted Morse potential
\begin{equation}
U(r)=\varepsilon\big[e^{-2\alpha(r-\ell_{ij})}-2e^{-\alpha(r-\ell_{ij})} +c_{\rm sh} \big] ,\ \ \    r \leq r_c
\label{eq_U}
\end{equation}
where $\ell_{ij}$ is the average diameter of particles $i$ and $j$; the interaction strength is $\varepsilon$ and $\alpha$ sets the range; the cutoff parameter is $r_c=1.4 \ell_{ij}$, and we truncate so that $U(r)=0$ for $r>r_c$; the constant shift $c_{\rm sh}$ is chosen so that $U$ is continuous at $r= r_c$.
We consider $N$ particles in a cubic box of volume $L^3$, with periodic boundaries.

The overdamped colloid motion is modeled through Langevin dynamics with a large friction constant.  Particles have mass $m$ and they evolve with Langevin dynamics so the position $\bm{r}_i$ of particle $i$ obeys $d\bm{r}_i/dt=\bm{v}_i$, and 
\begin{equation}
m\frac{d\bm{v}_i}{dt}=-\nabla_i V -\lambda (\bm{v}_i - \bm{u}^{\rm aff}_i) +\sqrt{2\lambda k_B T}\bm{\xi}_i
\label{SIeq_lan}
\end{equation}
where $V= \sum_{1\leq i < j\leq N} U(|\bm{r}_i-\bm{r}_j|)$ is the total potential energy, $\lambda$ is the friction constant, $\bm{u}^{\rm aff}_i$ is the local velocity of the (implicit) solvent,  and $\bm{\xi}$ is a standard Gaussian white noise.   (In the absence of shear flow $\bm{u}^{\rm aff}_i=0$, the sheared case is discussed below.)  The velocity damping time is $\tau_d=m/\lambda$.  All simulations are performed in LAMMPS~\cite{LAMMPS22}.

To avoid crystallization we consider a size polydisperse system. We have taken $7$ types of particles with diameters equally spaced between $0.76\bar\ell$ and $1.24\bar\ell$, 
with relative concentrations $$[0.0062, 0.0606, 0.2417, 
0.3829, 0.2417,0.0606, 0.0062]$$ to mimic a Gaussian distribution of diameters with $8\%$ polydispersity.

We work with non-dimensionalized parameters throughout.
As in~\cite{bhaumik2025yielding}, we estimate the critical interaction strength for spinodal decomposition as $\varepsilon^*\approx3.1 k_{\rm B}T$~\cite{noro2000}.  We report interaction strengths relative to this boundary, we focus on two state points with $\varepsilon = 4.5 k_{\rm B}T = 1.45\varepsilon^*$ and $\varepsilon = 10 k_{\rm B}T = 3.22\varepsilon^*$.  Our unit of distance is the mean particle diameter $\bar\ell$ and the unit of time is $\tausim=(m\bar\ell^2/k_{\rm B} T)^{1/2}=1$.
The colloid volume fraction is $\phi=\pi N\overline{\ell^3}/(6L^3)$.
The dimensionless interaction  range is 
$\alpha_0=\alpha \bar\ell$, we set $\alpha_0=33$ throughout.
The non-dimensionalised friction is $\lambda_0 = \lambda \tausim/m$.  We take $\lambda_0=10$ to mimic overdamped dynamics, increased $\lambda_0$ leads to similar results but the computational cost is higher.  The integration time step is $\Delta t=0.001 \tausim$.
A natural time scale for colloid motion is the Brownian time $\tau_B  = \bar\ell^2\lambda/(24 k_BT)$ which is the typical time for an overdamped free particle to diffuse its radius.  For the parameters chosen here $\tau_B = 0.417\tausim$.

This simulation method is computationally efficient and has been shown to capture the essential features  of depletion gels~\cite{zia2014micro, landrum2016delayed}, including quantitative comparisons with experiment~\cite{patrick2008direct, razali2017effects}.  As such, it complements modelling approaches that focus on the gels' network structure~\cite{colombo2014stress} but do not resolve the internal structure of the gel strands.
On the other hand, our model neglects hydrodynamic interactions, which do affect some aspects of gel structure~\cite{royall2015probing,graaf2018hydro}.

In the following, we have not attempted a quantitative matching of the parameters of simulation and experiment, because the implementation of the cyclic shear is necessarily different (not least that the simulation uses periodic boundary conditions, and an idealized thermostat that has to absorb the dissipated heat during the shear but has not been parameterised for the experiment).  Despite these differences, we will show that the simulations and experiments exhibit the same qualitative behaviour.

\subsubsection{Simulations}

\paragraph{Initialisation:}
Simulations are initialised in random configurations at volume fraction $\phi=0.2$.  Gels are formed by simulating for a time $t_{\rm w}$, during which spinodal decomposition occurs. Unless otherwise stated we simulate $t_{\rm w}=3\times 10^4\tausim$ ($\approx 7\times 10^4\tau_B$) and system size of $N=10^4$ particles; all results are averaged over many independent runs (typically 50), to enable statistically robust conclusions.

\begin{figure}[t]
\centering
\includegraphics[width=1\linewidth]{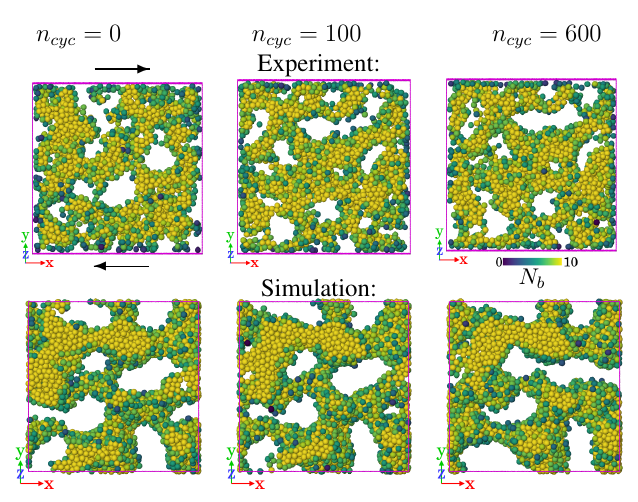}
    \caption{{\bf Visualization of gel under cyclic shear (experiment and simulation).} 
    (Top)~Rendering of a slice through an experimental gel, to visualize the strands and pores after different strain cycles $n_{cyc}=0,100,600$ {for the gel with $c_{\rm p}=1.25c_{\rm p}^*$}. Arrows indicate the geometry of the (simple) shear.  The region shown has volume $L\times L\times Z$ with $Z=8\ell$. (Bottom)~Similar slices based on simulation data for $\eps_0=1.5\eps^*$, $\gmax=0.04$. Particles are coloured by their coordination number, according to the colorbar.}
    \label{fig_expsim_slice}
\end{figure}

\paragraph{Cyclic shear:} 
After gel preparation we perform strain-controlled cyclic shear at finite temperatures and strain rates. The strain $\gamma_{xy}=\gamma(t)$ of the system is varied cyclically with a triangular wave with amplitude $\gamma_{\rm max}$ and rate $|\dot{\gamma}|=0.01\tausim^{-1}$.   This cycle is repeated many times.  (The simulated shear rates are significantly slower than those of the experiment, to ensure that the thermostat can easily absorb the energy that is injected by this external work.  We would expect qualitatively similar behaviour for large shear rates.)  We compute observable quantities at the stroboscopic configuration ($\gamma=0$) after each cycle of strain and study the evolution of these quantities as a function of number of cycles  $\ncyc$.  

We use Lees-Edwards boundary conditions. The solvent velocity $\bm{u}^{\rm aff}_i$ in \eqref{SIeq_lan} points in the $x$ direction, with magnitude $y_i \dot\gamma$, where $y_i$ denotes the $y$-component of particle $i$.

\paragraph{Creep Simulation:}
Gels are non-equilibrium states and the cyclic shear affects their structure.  After shearing, we performed 
stress-controlled creep dynamics to probe gels' yielding behaviour.  The procedure follows that of~\cite{bhaumik2025yielding}, a non-dimensionalized stress $\sigma_0 = \sigma \bar\ell^3/(k_B T)$ is maintained through a feedback control scheme that is implemented~\cite{VezirovSM2015,CabrioluSM19} as
\begin{equation}
\partial_t 
\dot{\gamma}=B[\sigma_0-\sigma_{xy}(t)]
\label{SIeq_rate}
\end{equation}
where $\dot\gamma$ is the shear rate and $\sigma_{xy}$ is the observed (non-dimensional) shear stress [measured from the virial]; also $B$  is the damping parameter determining how quickly the applied stress relaxes to its imposed value.  We take $B=0.01\tausim^{-2}$ as in~\cite{bhaumik2025yielding}, which ensures that the stress is imposed accurately, at a reasonable computational cost.
Similar to the cyclic shear, we used Lees-Edwards boundary conditions with affine flow in the $x$-direction.

\section{Results -- Gel structure\\ (experiment and simulation)}
\label{sec:results-structure}

\subsection{Gel visualisation}

Figure \ref{fig_expsim_slice}(top) shows slices through the experimental system.  (Specifically, we visualise particles in a region of size $L\times L\times Z$ with $Z=8\ell$, for various $n_{\rm cyc}$.)  The cyclic shear changes the gel structure: For $n_{cyc}=0$ the pores inside the gel are almost isotropic, but they become elongated as $n_{cyc}$ increases. 
Figure \ref{fig_expsim_slice}(bottom) presents similar slices from numerical simulation.  As noted abvoe, we have not attempted to match the parameters quantitatively, in particular the waiting times before start of shear are significantly longer in the simulation (leading to thicker gel strands).   Even so, the shearing leads to similar structural changes in both samples.

In the following subsections we analyse the results in detail for both experiment and simulation.  We mainly focus on the structural change at different length scales due to cyclic shear deformation.  From the simulations, we also show in Sec.~\ref{sec:results-rheo} that the cyclic shear hardens the gel.

\subsection{Experiment}

\subsubsection{Microscopic structure} 

To probe microscopic structural changes, we measure several local quantities, which mirror those measured in the simulation study of~\cite{bhaumik2025yielding}.  Specifically, we consider the coordination number $N_b$, the two-fold bond-orientation parameter $q_2$, and the average number of different types of clusters in which particles participate, as obtained form the TCC~\cite{malins2013tcc}. Particles within a distance of $1.6\ell $ of particle $i$ are defined as its neighbours. (This identification method for neighbours is maintained for all the experimental data analysis.)  The bond orientational order parameter $q_l$ is calculated following~\cite{SteinhardtPRB83}:
\begin{align}
    q_{l}(i)=\sqrt{\frac{4\pi}{2l+1}\sum_{m=-l}^{l}\left|q_{lm}(i)\right|^2},\\
    q_{lm}(i)=\frac{1}{N_b(i)}\sum_{j=1}^{N_b(i)}Y_{lm}(\mathbf{r}_{ij}),
\end{align}
where $Y_{lm}$ are spherical harmonics and $N_b(i)$ is the number of neighbours of reference particle $i$. We consider the case $l=2$ so that the quantity $q_2$ measures the stretching of a bond between two particles. 

\begin{figure}[t]
    \centering
    \includegraphics[width=1\linewidth]{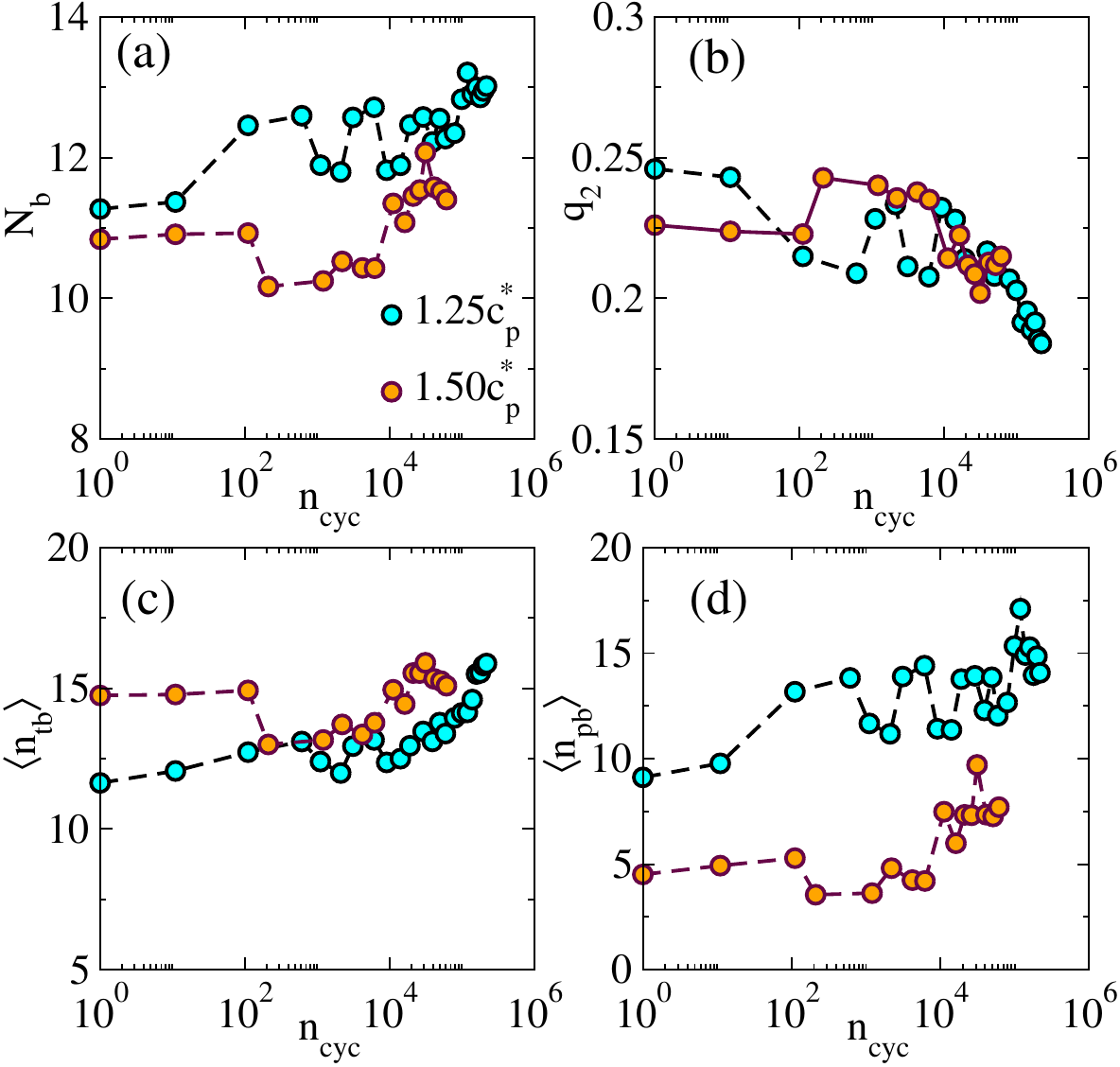}
    \caption{{\bf Microscopic structural analysis of sheared gel (experiment)}  (a) Coordination number $N_b$ as a function of the number of shear cycles $n_{\rm cyc}$. (b) Two-fold bond orientation order parameter. (c,d) TCC analysis, showing the number of trigonal bipyramids and pentagonal bipyramids in which the particles participate.
    See Fig.~\ref{fig_sim_microstruc} for a similar analysis of simulation data.}
    \label{fig_exp_microstruc}
\end{figure}

In Fig.~\ref{fig_exp_microstruc}, we show the evolution of these quantities as the system is sheared for two different gel samples.  On increasing $\ncyc$, the average coordination number $N_b$ increases [Fig.~\ref{fig_exp_microstruc}(a)].  This is due to a coarsening effect whereby the shearing increases the thickness of the gel strands. Fig.~\ref{fig_exp_microstruc}(b) shows that $q_2$ has a a decreasing trend with $\ncyc$. This $q_2$ is associated with stretching of interparticle bonds: one tends to find larger values in thinner strands, and smaller values in thicker strands where the structure resembles the bulk.  Hence the decreasing $q_2$ is consistent with the increasing $N_b$ due to coarsening. 

Figs. \ref{fig_exp_microstruc}(c,d) show the variation of the number of trigonal bipyramids $\langle n_{tb} \rangle$ and pentagonal bipyramids $\langle n_{pb} \rangle$ in which a particle participates as a function of strain cycles. These TCC structures contain $5$ and $7$ particles respectively, and are sensitive to details of the packing. Both the quantities show an increasing trend with $\ncyc$, indicating that the system evolves downhill in the energy landscape as the shear-induced coarsening (and aging) takes place.

\begin{figure}[t]
    \centering
    \includegraphics[width=1\linewidth]{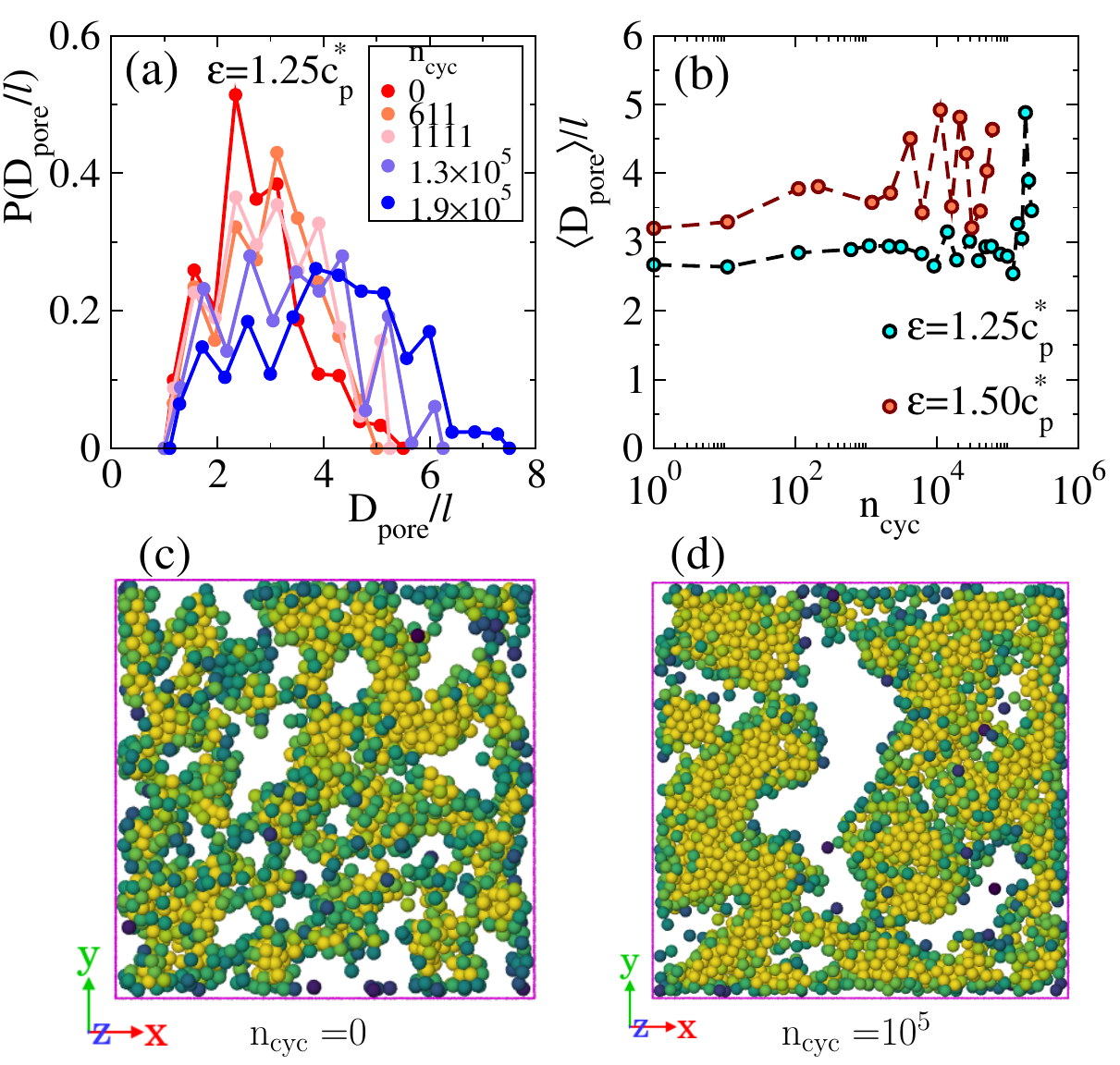}
    \caption{{\bf Mesoscopic (pore-size) analysis of sheared gel (experiment)}.   (a) Distribution of pore diameter for different numbers of strain cycles for the gel with $c_p=1.25 c_p^*$.
    (b)~Average pore diameter against the number of cycles for two different gels.  (c) Snapshot of a slice through gel before shearing  ($c_p=1.25 c_p^*$).  (d) Snapshot of the same gel after $10^5$ shear cycles. 
    Color coding in (c,d) is the same as Fig.~\ref{fig_expsim_slice}.  
    All pore-sizes $D_{\rm pore}$ are measured in units of colloid diameter $\ell$.
    See Fig.~\ref{fig_sim_mesostruc} for a similar analysis of simulation data.
}
\label{fig_exp_mesostruc}
\end{figure}

\subsubsection{Mesoscopic structure} 

We now consider the mesoscopic structure of the gel by measuring its pore size distribution.  We measure the distribution of pore sizes in the gel as defined in~\cite{GubbinsLang99}, the detailed procedure is described in~\cite{bhaumik2025yielding}:
We select a random point in the gel and we find the largest possible sphere that encompasses that point, without overlapping with any colloidal particles.  We identify the pore size $D_{\rm pore}$ as the diameter of this largest sphere. By repeating this process for many random points we obtain the distribution of pore sizes. 

Fig. \ref{fig_exp_mesostruc}(a) shows the distribution of the pore sizes $D_{\rm pore}$, as the shear accumulates, for the gel with $\eps=1.25\eps^*$. 
The distribution shifts towards larger pores as the shear $\ncyc$ increases. The maximal $D_{\rm pore}$ also has an increasing trend with $\ncyc$.  The associated mean pore size [Fig. \ref{fig_exp_mesostruc}(b)] has a mild initial increase before growing much more suddenly.  We find that the isotropic pores of the unsheared gel structure become anisotropic on shearing [recall Fig.~\ref{fig_expsim_slice}]; also, pores can grow very large after extensive shearing, see Fig. \ref{fig_exp_mesostruc}(c,d) for $\ncyc=0,10^5$ respectively. 
For the sample {with $\eps=1.5\eps^*$}, the data of Fig. \ref{fig_exp_mesostruc}(b) are consistent with an increasing trend for the mean pore size, but the results are subject to large fluctuations because the data contains a small number of large pores which strongly affect the estimated average.  (Recall that the location of the visualisation box is not constant in time, so the large changes as a function of $\ncyc$ do not represent large \emph{local} changes.)

\subsection{Numerical Simulation}

We now analyse results of numerical simulations, and compare with experiment. We consider two state points with different interaction strengths $\eps=1.45\eps^*, 3.22\eps^*$.  At the microscopic level it has been shown previously for this system~\cite{bhaumik2025yielding} that increasing $\eps$ during gel preparation leads to strands of reduced thickness with smaller $N_b$ and larger $q_2$.  Here we analyse the effect of the cyclic shear.

\begin{figure}
\centering
 \includegraphics[width=1\linewidth]{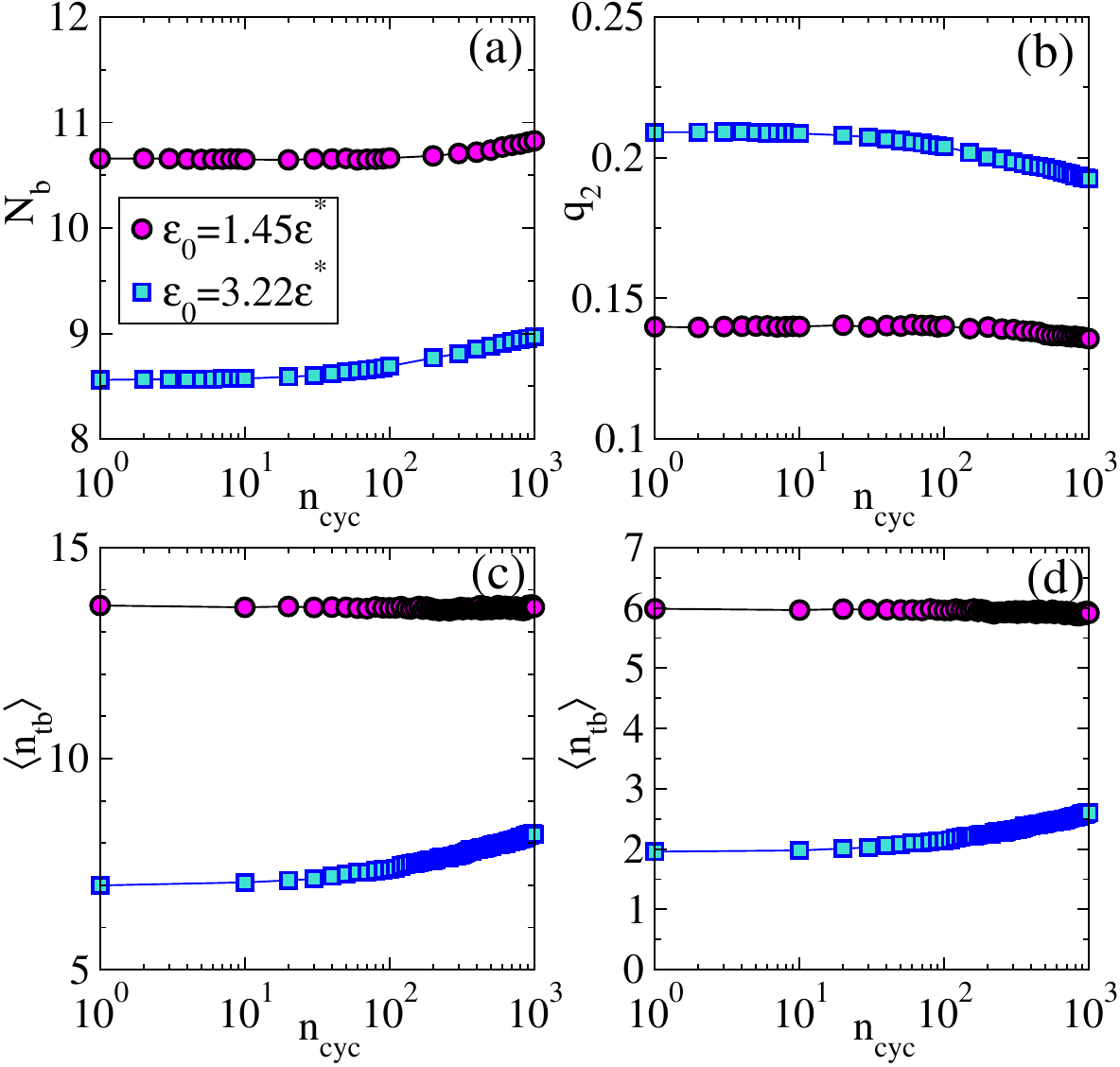}
    \caption{ {\bf Microscopic structural analysis of sheared gel (simulation).}  (a) Coordination number $N_b$ for increasing shear $\ncyc$. (b) Bond orientation order parameter $q_2$. (c,d)~TCC analysis showing the number of trigonal bipyramids and pentagonal bipyramids in which the particles participate.  (These results may be compared with Fig.~\ref{fig_exp_microstruc}.)  }
    \label{fig_sim_microstruc}
\end{figure}

To analyse \emph{microscopic} structure, particles within the interaction range $r_c$ of particle $i$ are defined as its neighbours for the evaluation of the co-ordination number $N_b$.  (This method of identifying neighbours is maintained throughout the analysis of numerical simulations.)  
We present the variation of $N_b$ and $q_2$ with the number of strain cycles in Fig. \ref{fig_sim_microstruc}(a,b).  The increasing behaviour of $N_b$ and a decreasing trend in $q_2$ with $n_{cyc}$ mirror the experimental observations of Fig.~\ref{fig_exp_microstruc}. In Fig. \ref{fig_sim_microstruc}(c) and (d) we present the data for $\langle n_{tb} \rangle$ and $\langle n_{pb} \rangle$ obtained from TCC analysis.  For $\eps=3.22\eps^*$,we see a significant increase in the value of $\langle n_{tb} \rangle$ and $\langle n_{pb} \rangle$  mirroring experimental data. 
For $\eps=1.45\eps^*$, the change in these data are negligible, despite the increasing co-ordination number.  This may indicate that the microscopic structure inside the gel strands is close to that of the (metastable) colloidal liquid, 
due to the long annealing times for the gel, and the relatively weak bonds.  Then the agitation by shearing has little effect on the structure, although there is some coarsening of the strands, which increases $N_b$.

\begin{figure}
     \includegraphics[width=1\linewidth]{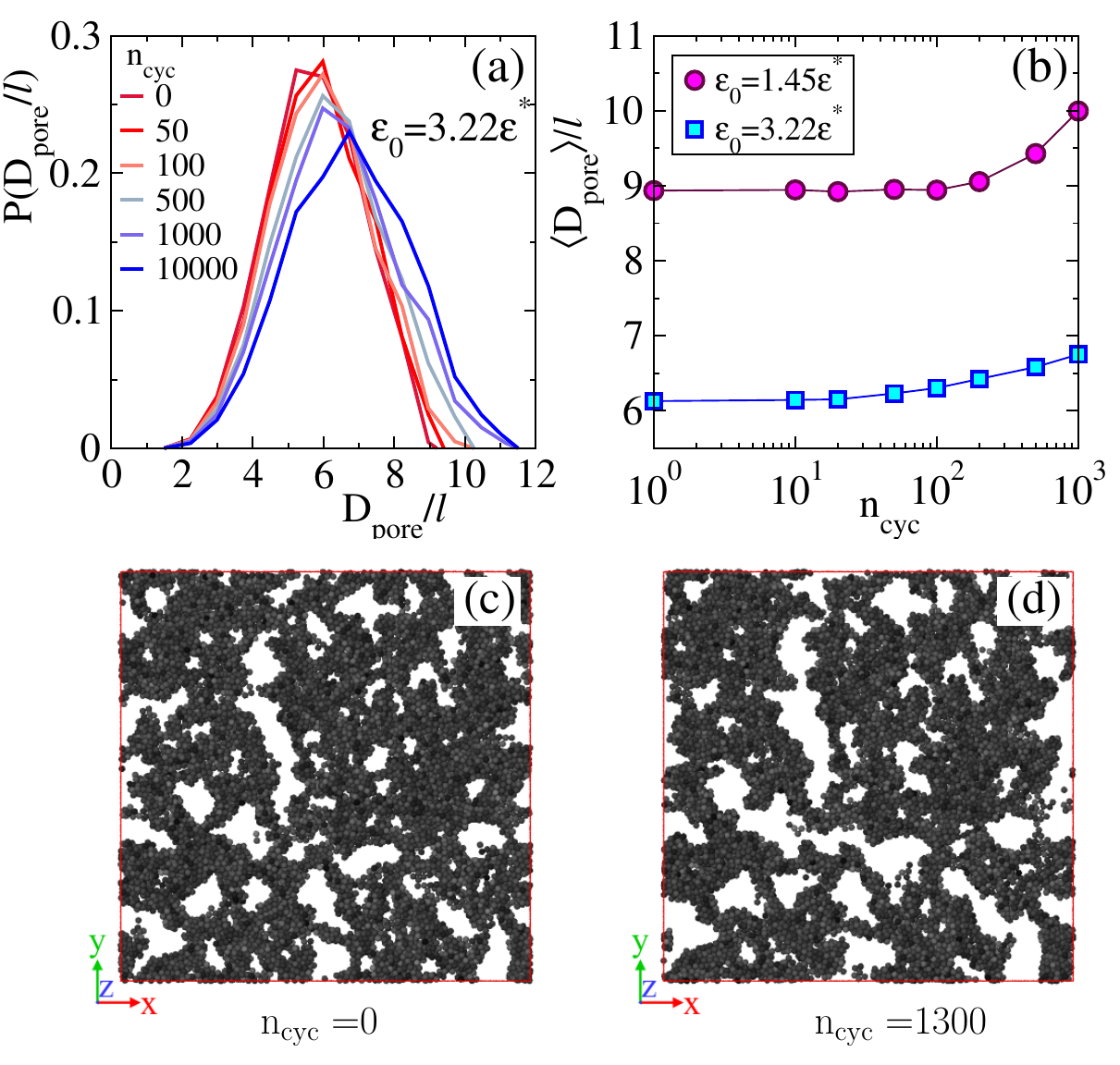}
    \caption{{\bf Mesoscopic (pore-size) analysis of sheared gel (simulation).} (a) Distribution of pore diameter for different numbers of strain cycles. (b) Average pore diameter against the number of cycles. 
    (c) Snapshot of a slice through the gel, before shearing (interaction strength $\eps=3.22\eps^*$).  (d) The same gel after $\ncyc=1300$ shear cycles.  (These results may be compared with Fig.~\ref{fig_exp_mesostruc}.) }
    \label{fig_sim_mesostruc}
\end{figure}

Next, we investigate the \emph{mesoscopic} structure of the system by analyzing the pore size of the gel. As above, we extract the distribution of the pore size $D_{\rm pore}$ and we extract its mean value.
Fig. \ref{fig_sim_mesostruc}(a) shows results for increasing $n_{cyc}$ in a gel with $\eps=3.22\eps^*$. As $n_{cyc}$ increases, the distribution shifts to the right with a larger mean. In Fig. \ref{fig_sim_mesostruc}(b) illustrates the increasing behaviour of average poresize with $n_{cyc}$ for $(\eps/\eps^*)=1.45,3.22$. To provide a visual representation we present snapshots of slices through the gel in Fig. \ref{fig_sim_mesostruc}(c,d) for $n_{cyc}=0$ and $n_{cyc}=1300$, respectively. Initially, at $n_{cyc}=0$, the system exhibits a smaller pore size. However, after multiple applied cycles, the pores elongate, contributing to an overall increase in the average pore size.

\subsection{Summary}

We find a coherent picture of the microscopic and mesoscopic structural changes in these cyclically sheared colloidal gels, both in the experiment and numerical simulations. The main effect is that shearing leads to a coarsening effect, where colloids' mean co-ordination number increases, as does the pore size of the gel.  This can be rationalised in a simple way by the idea that the shearing injects energy and accelerates the random bond-breaking processes that are anyway occurring due to thermal fluctuations.  In this way, it accelerates the natural aging/coarsening dynamics of the gel.  However, the shearing procedure also introduces anisotropy to the system, which is apparent in the shape of the pores in the gel [Fig.~\ref{fig_expsim_slice}].

Overall, the simulation model successfully captures the key features of structural changes across different length scales resulting from the repetitive mechanical deformation of the gel, including the microscopic structure of the arms.  This complements previous results for effects of shear on the network topology, in which the internal structure of strands was not modelled in detail~\cite{colombo2014stress,Bouzid2018langmuir}.
In the next Section, we use numerical simulations to investigate the {mechanical} 
response of the gel in more detail.

\section{Results -- Mechanical Properties 
(simulation)}
\label{sec:results-rheo}

\begin{figure}[t]
    \centering
    \includegraphics[width=1\linewidth]{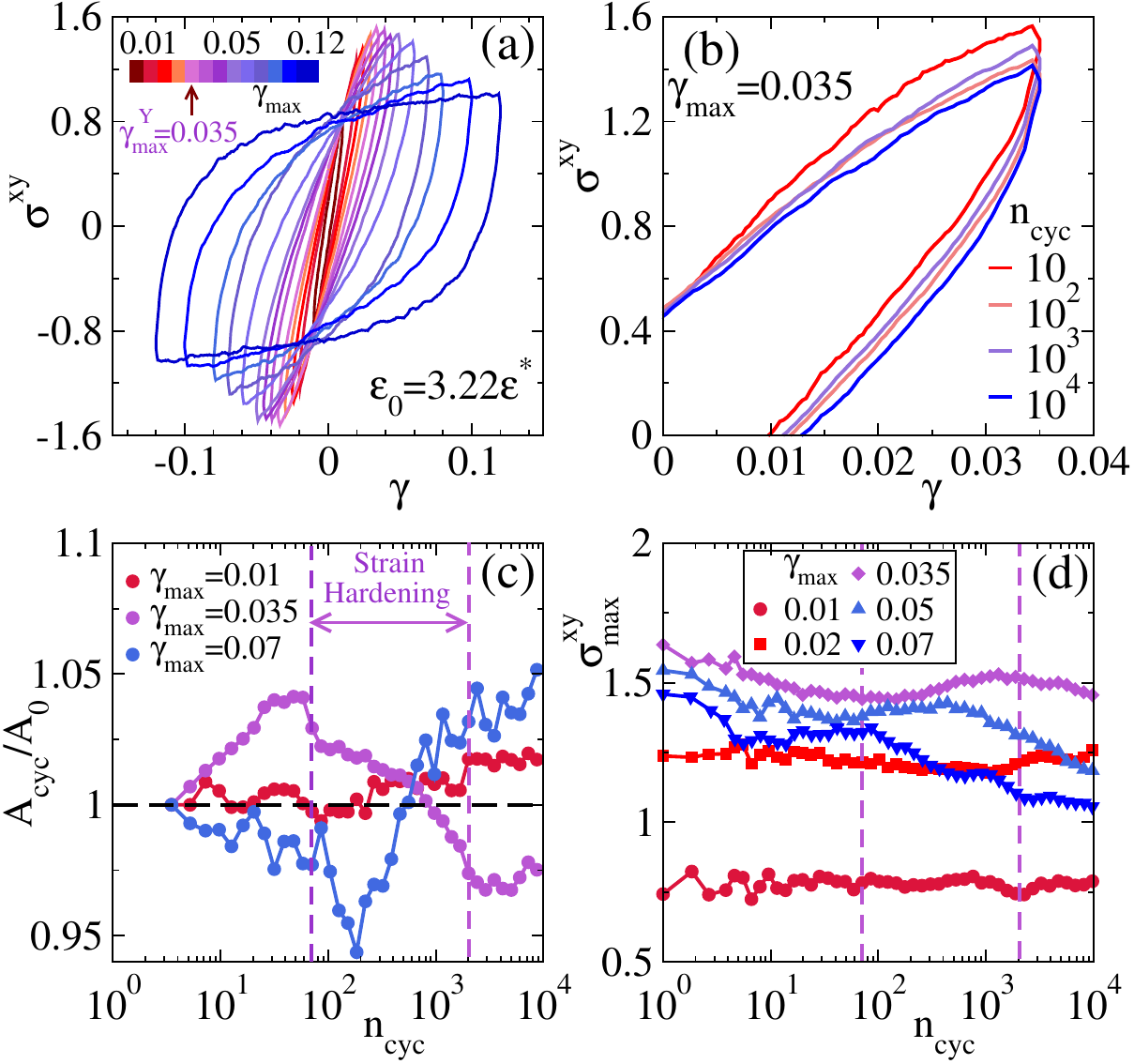}
 \caption{\textbf{Cyclic shear for a simulated gel}, interaction strength $\eps=3.22\varepsilon^*$. (a)~Average stress-strain curve over a strain cycle for several values of $\gamma_{max}$ ranging from $0.02$ to $0.12$.  (Data are averaged over the final 200 cycles, obtained from simulations of $10^4$ cycles.)
 (b)~Zoomed view of stress-strain curves for $\gamma_{max}=0.035$ at different values of $n_{cyc}$. (c) The area  enclosed by the stress-strain curve as a function of $n_{cyc}$ (normalized with the value of initial cycle $A_0$), for three different values of $\gamma_{max}$. (d)~Maximum stress $\sigma^{xy}_{max}$ at $\gamma_{xy}=\gamma_{max}$ in a strain cycle against $n_{cyc}$ for several $\gamma_{max}$.}
 \label{fig_sim_hardening}
\end{figure}

So far, we considered cyclic shear of fixed amplitude $\gmax=0.04$ in both experiment and simulation, and we analysed structural properties of the gel as a function of the number of cycles.  In this Section we exploit the ability of simulation to measure the time-dependent stress during cyclic shear.  We also explore the dependence on the strain amplitude, revealing an interesting regime where the gel hardens under shearing.

\subsection{Strain hardening before failure} 

\subsubsection{Hysteresis loops}

We consider gels with $\eps=3.22\eps^*$, as visualised in Fig.~\ref{fig_sim_mesostruc}.
Simulation results are shown in Fig.~\ref{fig_sim_hardening} for a range of $\gmax$.  Specifically, Fig.~\ref{fig_sim_hardening}(a) shows stress-strain curves, obtained from long simulations ($\ncyc=10^4$), averaged over the final 200 cycles.  For very small $\gmax$ and small shear rates $\dot\gamma$ one expects an elastic response with $\sigma_{xy} \propto \gamma$ and no hysteresis loop.  We find here that some hysteresis is present even for $\gmax=0.01$, which we attribute to the fact that the rate $|\dot\gamma|=0.01\tausim^{-1}$ is not extremely small.  

On increasing the amplitude $\gmax$, the area of the hysteresis loop increases; the magnitude of the shear stress also depends non-trivially on $\gmax$.  This non-trivial dependence is associated with increasing dissipation, the onset of plastic deformation, and yielding.  To identify the transition between elastic and plastic flow, we write $\sigma_{\rm max}^{xy}$ for the maximal stress during the cycle: this has a non-monotonic dependence on $\gmax$~\cite{Leishangthem2017,bhaumik2021role}.  There is no sharp yielding transition but we follow~\cite{Leishangthem2017} in identifying the maximal value of $\sigma_{\rm max}^{xy}$ with the transition to plastic flow; the associated shear amplitude is $\gmax^{\rm Y}$.  From Fig.~\ref{fig_sim_hardening}(a) we identify $\gmax^{\rm Y}=0.0350\pm 0.0005$.

In Fig. \ref{fig_sim_hardening}(b) we show the evolution of the stress-strain loop as the number of shear cycles increases, for $\gmax=0.035$, close to the yield strain.\footnote{Each loop is averaged over $10$ cycles, they are not perfectly symmetric under inversion through the origin because the system is evolving structurally throughout the shearing process.} Close inspection of the four curves shows that the dependence on $\ncyc$ is non-monotonic.  Fig.~\ref{fig_sim_hardening}(c) plots the area  of the loop (averaged over $10$ consecutive cycles) as a function of $\ncyc$, normalised by the value in the initial cycle.  For $\gmax=0.035$ the non-monotonic dependence is clear.  We also define  $\sigma_{\rm max}^{xy}$ as the maximal stress during the cycle.  This quantity is shown in  Fig.~\ref{fig_sim_hardening}(d), again showing non-monotonic dependence on $\ncyc$ for $\gmax=0.035$ (see also Fig. S9 of Ref.~\cite{bhaumik2021role} for an analogous effect in glasses).

These non-monotonic dependencies are associated with strain-hardening of the gel, as we now explain. 
Note first that for small strain amplitude $\gmax=0.01$, the area and the maximal stress hardly depend on $\ncyc$.  This indicates that the gel is mostly responding elastically and the repeated shear cycles have little effect on its structure.  For larger $\gmax$ the overall trend is that the loop area increases and the maximal stress $\sigma_{\rm max}^{xy}$ decreases.  This indicates that the gel's response includes plastic rearrangements that disrupt its structure and lead to extra dissipation (larger loop area); the disruption to the structure also increases the compliance (reduces $\sigma_{\rm max}^{xy}$).  However, for $\gmax=0.035$, the range labelled ``strain hardening'' in the figure is associated with a reduction in both the dissipation and the compliance.  
Similar effects are observed for other values of $\gmax$ although the signature is most pronounced for $\gmax\approx \gmax^{\rm Y}$.  Depending upon the strain amplitude, the region of $n_{\rm cyc}$ at which the strain hardening is observed is varied.  

\subsubsection{Dynamic moduli}

\begin{figure}[t]
    \centering
    \includegraphics[width=1\linewidth]{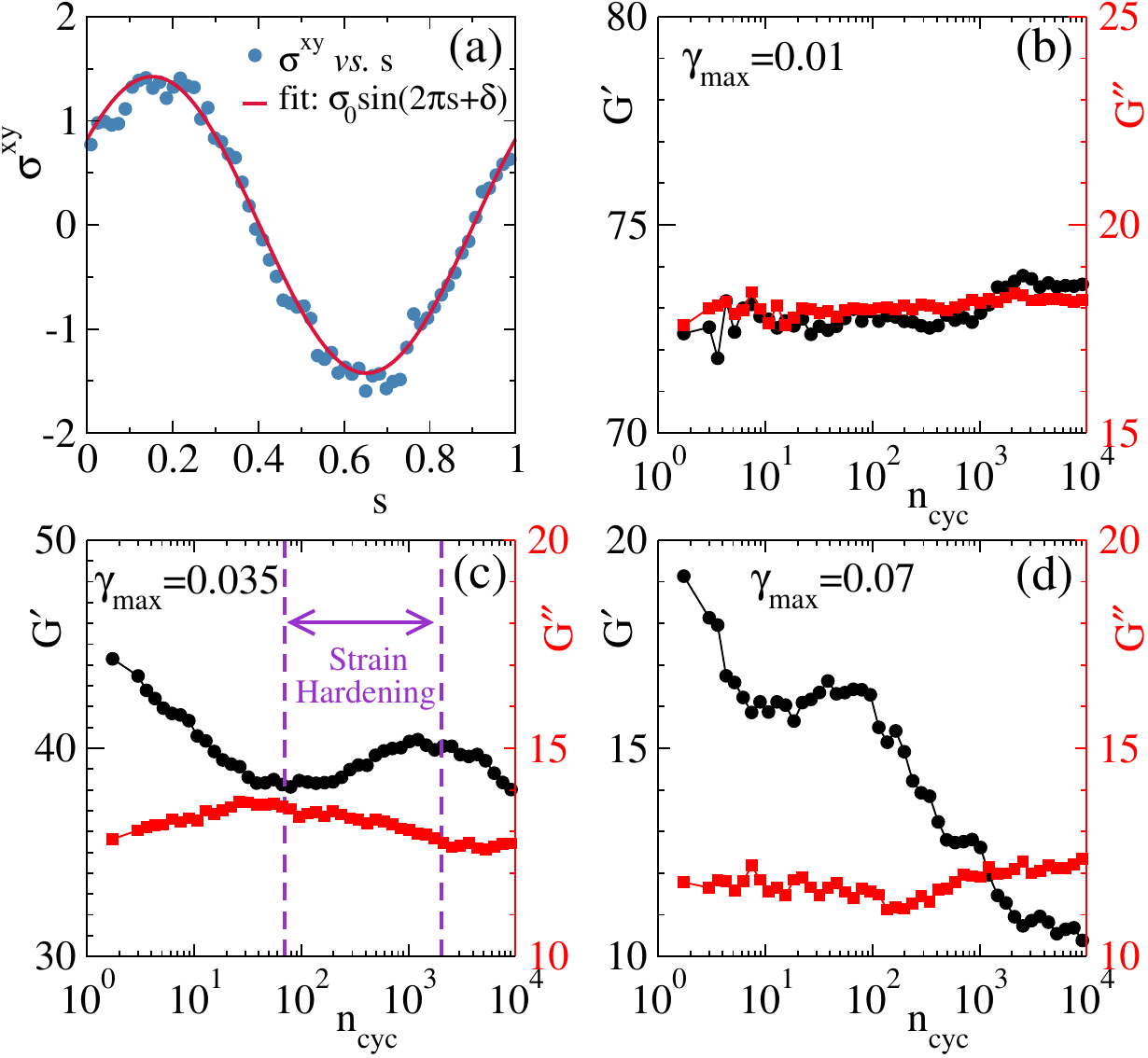}
 \caption{\textbf{Estimated shear moduli for a simulated gel}, based on shear protocols with different amplitudes (interaction strength $\eps_0=3.22\eps^*$).
 (a)~Stress $\sigma^{xy}$ as a function of rescaled time $s$. Solid line through the data points is a fit with $\sigma^{xy}=\sigma_0 \sin(2\pi s+\delta)$. Sample averaged storage modulus $G^\prime$ and loss modulus ($G^ {\prime \prime}$) as a function of strain cycle for (a) $\gamma_{max}=0.01$, (b) $\gamma_{max}=0.035$, and (c) $\gamma_{max}=0.07$. }
 \label{fig_sim_modulus}
\end{figure}

To probe further the strain-hardening phenomena, we investigate dynamic moduli (storage modulus $G^ \prime$, and loss modulus $G^{\prime \prime}$).  More specifically, we measure  ``nonlinear moduli'' which incorporate the dependence of the stress response on the frequency and amplitude of an oscillatory (strain-controlled) shear.  These quantities are familiar from studies of large amplitude oscillatory shear (LAOS)~\cite{HyunProgPolySc2011}.  
However, the standard theory is formulated for sinusoidal shear cycles, while our numerical data is obtained with a triangular (non-sinusoidal) waveform.  To estimate the moduli in this case, we follow~\cite{YehPRL2020} and reparameterise the time $t$ in terms of a variable $s=s(t)
$ such that
$\gamma(t)=\gamma_{\rm max}\sin(2\pi s(t))$. We denote the time taken for a single shear cycle as $\tau_{\rm shear}=\gamma_{\rm max}/(4\dot\gamma)$, also note that $s(\tau_{\rm shear})=1$ and define the corresponding angular frequency as  $\omega=(2\pi/\tau_{\rm shear})=(8\pi\dot\gamma/\gamma_{\rm max})$.

{The stress profile for any shear cycle can then be fitted as a function of $s$, as $\sigma(s) = \sigma_0 \sin(2\pi s+\delta)$ where the amplitude $\sigma_0$ and phase lag $\delta$ are fitting parameters.  
An example fit is shown in Fig. \ref{fig_sim_modulus}(a). 
This allows nonlinear moduli $G^\prime$ and $G^{\prime\prime}$ to be extracted from the formula
\begin{equation}
G^\prime + iG^{\prime\prime} =\frac{\sigma_0}{\gamma_{\rm  max}} [\cos(\delta)+i \sin(\delta)]
\end{equation}
where $\sigma_0,\delta$ are obtained from the fit, while the amplitude $\gamma_{\max}$ is set by the underlying shear cycle.  Since the gel is hardening as the shear accumulates, we perform this fitting separately for each individual cycle.  We average the resulting $\sigma_0,\delta$ over many independent samples.\footnote{We verified that similar results are obtained by averaging the time-dependent stress over the samples and then fitting the average.}

The resulting moduli depend on $(\omega,\gamma_{\rm max},n_{\rm cyc})$; recall that $\dot\gamma$ is fixed throughout this study so we study $G',G''$ as functions of $(n_{\rm cyc},\gamma_{\rm max}$). Results are shown in Fig. \ref{fig_sim_modulus}(b-d) for different values of strain amplitude.  In each case we measure how the moduli change as the oscillatory shear accumulates (increasing $n_{\rm cyc}$).  We discuss the different amplitudes in turn.

 For small amplitude $\gamma_{\rm max}=0.01$ [Fig. \ref{fig_sim_modulus}(b)], both $G^\prime$ and $G^{\prime\prime}$ remain almost constant.  
 For larger amplitude  $\gamma_{\rm max}=0.035$ (close to the yield strain), Fig. \ref{fig_sim_modulus}(c), shows more interesting behaviour, in that both $G^\prime$ and $G^{\prime\prime}$ are non-monotonic functions of the number of cycles $n_{\rm cyc}$. An initial decrease in $ G^ \prime$ and a corresponding increase in $G^{\prime\prime}$ indicate gel softening. At intermediate cycle number, however $G^\prime$ begins to rise while $G^{\prime\prime}$ decreases, indicating hardening of the gel. Notably, this transition region aligns with our earlier observations -- specifically, the reduction in the stress-strain loop area and the increase maximum stress as shown in Fig. \ref{fig_sim_hardening}(c,d). 
 
 For a yet larger strain amplitude $\gamma_{max}=0.07$ [Fig. \ref{fig_sim_modulus}(d)], a similar hardening behavior is observed, but it is restricted to a narrower range of cycles ($n_{\rm cyc}=10$-$100$). Beyond this regime the gel exhibits liquid-like characteristics, as indicated by  $G^\prime< G^{\prime \prime}$ for $n_{\rm cyc}>10^3$.
Taken together, these results complement previous works that demonstrated strain-hardening in other gel models, which had different microscopic interactions (and where thermal fluctuations were neglected)~\cite{colombo2014stress,Bouzid2018langmuir}.

\subsection{Delayed failure time of cyclically sheared gel}

Strain-hardening behavior is evident from Fig.~\ref{fig_sim_hardening}, the physical interpretation is that weak strands in the gel are broken by the shearing and the displaced particles are absorbed into the remaining strands, leading to a stronger (less compliant) gel.  One may expect that the resulting gels are also more stable: we now demonstrate that this is indeed the case.

  \begin{figure}[t]
    \centering 
    \includegraphics[width=1\linewidth]{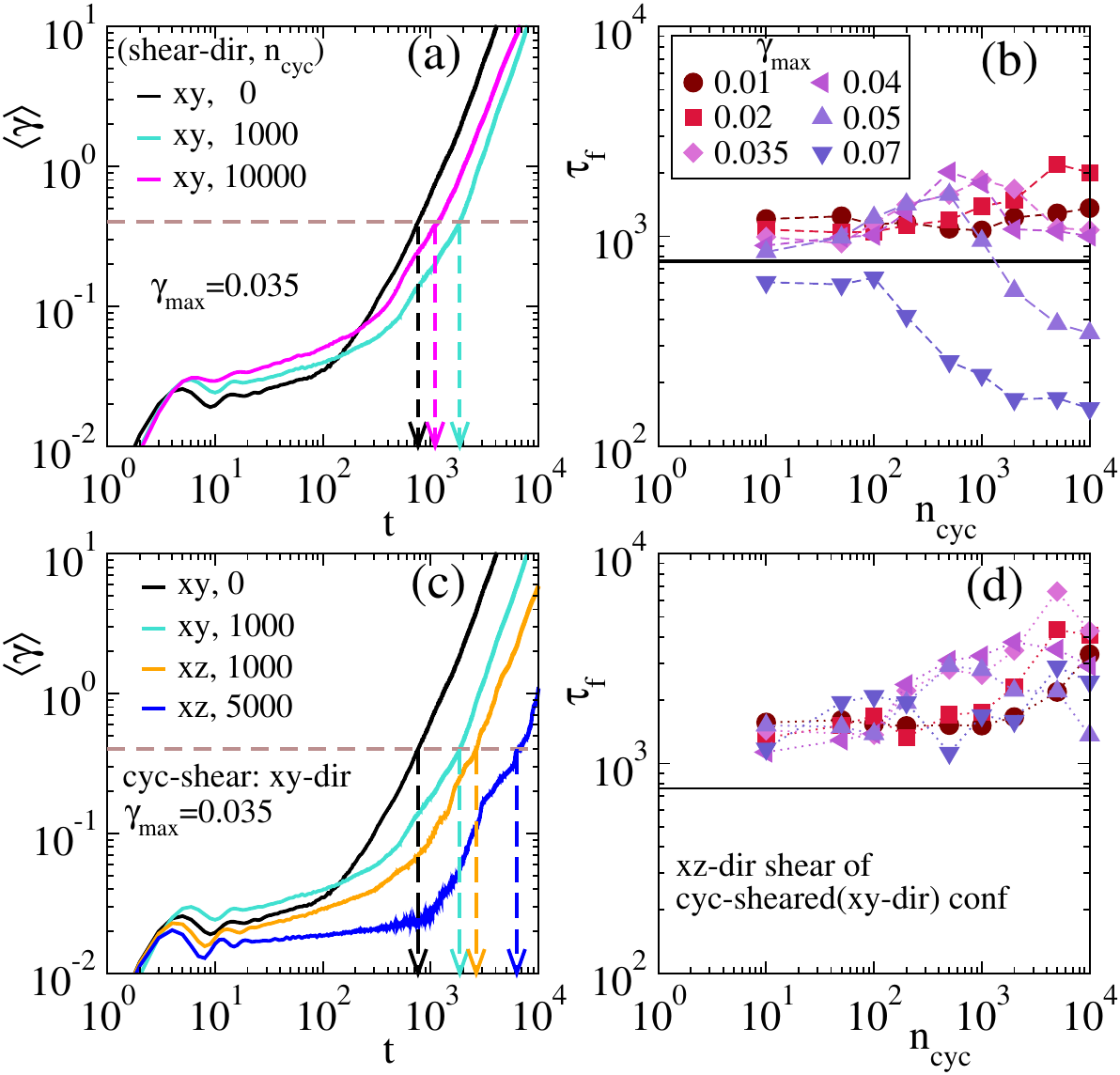}
    \caption{{\bf Creep Simulation}, interaction strength $\eps_0=3.22\eps^*$
     (a)~Average strain against time for different values of $n_{\rm cyc}=0,10^3,10^4$. The dashed horizontal line represents $\gamma=0.4$, which is taken as the critical value of strain beyond which system starts to flow. Vertical dashed lines indicate the failure time $\tau_{\rm f}$ for each case. (b) Failure time $\tau_{\rm f}$ as a function of $n_{\rm cyc}$ for different $\gamma_{\rm max}$. (c) Strain against time for shear applied to different directions and (d) the corresponding failure time as a function of $n_{\rm cyc}$.}    \label{fig_sim_creep}
\end{figure}

To this end, we simulated creep dynamics at constant stress, following~\cite{bhaumik2025yielding}.
The initial conditions for this procedure are  gel configurations that have been cyclically sheared for $\ncyc$ strain cycles. We then impose a constant (non-dimensionalised) shear stress $\sigma_0$ in the $xy$ plane, allowing flow along the $x$-direction with Lees-Edwards boundary conditions  (which is the same geometry used for the cyclic shear).  We take $\sigma_0=0.8$, this value is chosen such that the system undergoes creeping flow and then fails within our computational time window.\footnote{See Ref.~\cite{bhaumik2025yielding} for further discussion of the $\sigma_0$-dependence: we would expect yielding to be similar at smaller $\sigma_0$, if sufficiently long simulations were performed. The behaviour for larger $\sigma_0$ is different because the system yields almost immediately.}   

We measure the strain during this creep flow: its average behaviour is shown in Fig. \ref{fig_sim_creep}(a), for samples that have been through $n_{\rm cyc}=0,10^3,10^4$ shear cycles, with $\gamma_{\rm max}=0.035$.   The time-dependent strain has an initial elastic branch followed by a plateau,  creeping flow, and eventually failure.  The cyclic shear increases the height of the plateau: its value is comparable with $\gmax$ indicating that the structure of the gel has rearranged during cyclic shear so that it can accommodate this strain~\cite{MemoryRMP2019}.

We define the failure time $\tau_{\rm f}$ of the gel as the time for $\langle \gamma \rangle$ to reach a threshold $\gamma^*=0.4$, as explained in Ref.~\cite{bhaumik2025yielding}.  From Fig. \ref{fig_sim_creep}(a), the failure time is largest for $\ncyc=10^3$ cycles, after which it decreases again.  This is consistent with Fig.~\ref{fig_sim_hardening}, which indicates that the strain-hardening period ends at around $\ncyc=2\times 10^3$.  That is, strain hardening is also accompanied by an increased failure time.

We performed creep simulations for a range of $\gmax$ and $\ncyc$, from which we computed the failure times. Results are collated in Fig. \ref{fig_sim_creep}(b), which we compare with ``pristine'' gels ($\ncyc=0$).
For $\gmax=0.035$, strain hardening persists up to $\ncyc=10^3$, beyond which the material begins to soften again. This reinforces the conclusion of Fig. \ref{fig_sim_creep}(a).  
The same characteristic non-monotonicity is also observed for other cyclic shear amplitudes in the range $0.02\leq \gmax \
\leq 0.05$, and the ranges of $\ncyc$ over which this is observed mirror the ranges of strain hardening in Fig.~\ref{fig_sim_hardening}(d).  For small amplitude, $\gmax=0.01$, the  failure time for $n_{\rm cyc}=10$ is enhanced with respect to the pristine gel.  After this, $\tau_{\rm f}$ is weakly affected by further shearing.  This mirrors the small structural change due to such cyclic shear [recall Fig.~\ref{fig_sim_hardening}].
For large amplitude $\gmax=0.07$ the failure time is reduced by the cyclic shear, presumably because this significantly disrupts the gel network.

These moderate enhancements of $\tau_{\rm f}$ by cyclic shear (up to a factor of 3) are affected by two competing processes.  The agitation by shearing promotes coarsening of the gel structure, and activates relaxation processes that allow it to descend down the energy landscape, as occurs in physical aging.  This tends to increase {the failure time} $\tau_{\rm f}$.  At the same time, the repeated cyclic strain imprints an anisotropic memory on the gel structure: this may predispose it to flow along the direction of the shear, reducing $\tau_{\rm f}$.  (This process may be understood through the framework of fatigue failure~\cite{maity2024arxiv,BhowmikPRE2022}, where damage accumulates progressively until the system ultimately fails.).

To disentangle these effects, we simulated creeping flow with constant stress applied in the $xz$ direction {(the flow is still along $x$ but the gradient is now in $z$)}.  Results for $\gmax=0.035$ are shown in Fig. \ref{fig_sim_creep}(c).  Comparing with Fig.~\ref{fig_sim_creep}(a), one sees that the enhancements of $\tau_{\rm f}$ are much stronger when the creep flow is not in the same plane as the cyclic shear.  Our interpretation is that the increase of $\tau_{\rm f}$ due to coarsening and aging effects is dominating this response due to $xz$-stress, because memory effects due to $xy$-shear are less relevant in this geometry.  That is, changing the direction of the  shear stress in creep simulations separates the coarsening/shearing effects from the damage accumulation due to previous cyclic shear in the $xy$ plane.
 Surprisingly, 
the increase of $\tau_{\rm f}$ with $\ncyc$ continues even up to $n_{cyc}=5000$, at which time Fig.~\ref{fig_sim_hardening}(d) suggests that the system has entered the shear-softening regime.  
This result further illustrates the complex dependence of gel structure and response on its mechanical history.

 Fig. \ref{fig_sim_creep}(d) plots the failure times for creeping flow with constant $xz$-stress, showing that $\tau_{\rm f}$ is generically increased by shearing, 
  regardless of whether the system is in the strain-hardening or softening regime.  It seems that the softening effect at large $\ncyc$ in Fig.~\ref{fig_sim_hardening}(c,d) is an anisotropic effect, as may be expected from the mechanical protocol used.
Given the strong anisotropy of the response [visible as differences between Fig.~\ref{fig_sim_creep}(b,d)], it would be interesting to explore in more detail the effects of different cyclic shear protocols on failure. (For example, combining cyclic shears in more than one plane~\cite{CohenSMMemory2020} might generate more isotropic structures.)

\section{Discussion}
\label{sec:conc}

Colloidal gels are arrested far from equilibrium, so it is expected that their properties are affected by their mechanical history, including cyclic shear.  However, it remains challenging to predict these effects and exploit them for material design: materials' history-dependence is complex in general and there may be competing effects.  For example the shearing may break gel strands, but it also leads to coarsening and hence to thicker strands.  In this situation, particle-resolved experiments and simulations offer valuable opportunities to disentangle different behaviors and to understand their microscopic mechanisms.

We have made progress in this direction, showing in both experiments and simulations that shearing leads to coarsening of the gel network and to denser microscopic packing within the strands.  Consistent with recent work on glasses, this illustrates the general principle that shearing offers a mechanism for accelerating the progress of materials towards lower free-energy states.  We also showed that these structural changes also affect material responses like compliance and stability.

Our understanding of these effects will inform future work that aims to predict effects of mechanical processing, and use them to design material properties.  Theoretical insights would be very valuable in this endeavour, because the multi-scale structure of gels means that many colloidal particles have to be simulated (or tracked) in order to model the behaviour of a single gel strand, while much of the important physics is happening on larger length scales, via properties of the gel network.  This points towards the development of simplified meso-scale models, although there are significant couplings between microscopic structure and the responses of gel strands, which indicate that these theoretical models should not be \emph{too} simple.  In any case, we are optimistic that prediction of history-dependent properties can be improved beyond the current state-of-the-art, via a combined approach based on theory, experiment, and computer simulations.

\begin{acknowledgments}
We thank Kris Thijssen, Abraham Mauleon-Amieva, Rui Cheng, Malcolm Faers, and Srikanth Sastry for helpful discussions. This work was supported by the EPSRC through grants EP/T031247/1 (HB and RLJ) and EP/T031077/1 (CPR and TBL). CPR acknowledges the grant DiViNew from the Agence National de Recherche. JEH and CPR also acknowledge the European Research Council (ERC consolidator grant NANOPRS, project 617266).  HB thanks SUPRA (India) for support through grant SPR/2021/000382 during a part of the period when this work was performed.
\end{acknowledgments}

\bibliography{Bibliography.bib} 
\end{document}